\documentclass[12pt, a4paper]{article}

\usepackage{rotating, verbatim}
\usepackage{epsfig}
\usepackage{latexsym}
\usepackage{graphicx}
\usepackage{psfrag}
\usepackage{amsmath,amssymb}
\usepackage{amsfonts}
\usepackage{amssymb}
\usepackage{graphicx, rotating}
\usepackage{epstopdf}
\usepackage{epsfig}
\usepackage{latexsym}
\usepackage{graphicx}
\usepackage{color}
\usepackage{amsmath,amssymb}
\usepackage{cite}
\usepackage{slashed}
\usepackage{hyperref}
\usepackage{datetime}
\usepackage{verbatim}

\makeatletter
\def\section{\@startsection {section}{1}{\z@}{-3.5ex plus -1ex minus
 -.2ex}{2.3ex plus .2ex}{\large\bf}}
\def\subsection{\@startsection{subsection}{2}{\z@}{-3.25ex plus -1ex
minus -.2ex}{1.5ex plus .2ex}{\normalsize\bf}}
\makeatother
\makeatletter
\def\theequation{\arabic{section}.\arabic{equation}}

\@addtoreset{equation}{section}
\renewcommand{\theequation}{\thesection.\arabic{equation}}
\makeatother

\newcommand{\captionfonts}{\small}
\makeatletter  
\long\def\@makecaption#1#2{%
  \vskip\abovecaptionskip
  \sbox\@tempboxa{{\captionfonts #1: #2}}%
  \ifdim \wd\@tempboxa >\hsize
    {\captionfonts #1: #2\par}
  \else
    \hbox to\hsize{\hfil\box\@tempboxa\hfil}%
  \fi
  \vskip\belowcaptionskip}
\makeatother   

\catcode`@=11
\def\marginnote#1{}
\newcount\hour
\newcount\minute
\newtoks\amorpm
\hour=\time\divide\hour
by60
\minute=\time{\multiply\hour by60 \global\advance\minute
by-\hour}
\edef\standardtime{{\ifnum\hour<12 \global\amorpm={am}
\else\global\amorpm={pm}\advance\hour by-12 \fi
 \ifnum\hour=0
\hour=12 \fi
 \number\hour:\ifnum\minute<10
0\fi\number\minute\the\amorpm}}
\edef\militarytime{\number\hour:\ifnum\minute<10
0\fi\number\minute}
\def\draftlabel#1{{\@bsphack\if@filesw
{\let\thepage\relax
 \xdef\@gtempa{\write\@auxout{\string
\newlabel{#1}{{\@currentlabel}{\thepage}}}}}\@gtempa
 \if@nobreak
\ifvmode\nobreak\fi\fi\fi\@esphack}
\gdef\@eqnlabel{#1}}
\def\@eqnlabel{}
\def\@vacuum{}
\def\draftmarginnote#1{\marginpar{\raggedright\scriptsize\tt#1}}
\def\draft{\oddsidemargin
0.0truein
 \def\@oddfoot{\sl preliminary draft \hfil
\rm\thepage\hfil\sl\today\quad\militarytime}
 \let\@evenfoot\@oddfoot
\overfullrule 3pt
 \let\label=\draftlabel
\let\marginnote=\draftmarginnote
\def\@eqnnum{(\theequation)\rlap{\kern\marginparsep\tt\@eqnlabel}
\global\let\@eqnlabel\@vacuum}
}
\catcode`@=12
\def\bea{\begin{eqnarray}} \def\eea{\end{eqnarray}}
\def\be{\begin{eqnarray}} \def\ee{\end{eqnarray}} 

\begin{document}

\thispagestyle{empty}

 \begin{center}
 \small
 \hfill ANL-HEP-PR-12-47
 \hfill  BI-TP-2012-22
 \hfill EFI-12-15
 \hfill FERMILAB-PUB-12-375-T
 \hfill  UAB-FT-706 

 \begin{center}

\vspace{1.7cm}

{\Large\sc  MSSM Electroweak Baryogenesis  and LHC Data}

\end{center}

\vspace{1.4cm}

 { \bf M. Carena$^{\,a}$,  G. Nardini$^{\,b}$, 
 M. Quir\'os$^{\,c}$, C.E.M. Wagner$^{\,d,e}$}\\

 \vspace{1.2cm}

 ${}^a\!\!$
 {\em {Theoretical Physics Department, Fermilab, P.O. Box 500, Batavia, 
 IL 60510}}

${}^b\!\!$ {\em {Fakult\"at f\"ur Physik, Universit\"at Bielefeld,
    D-33615 Bielefeld, Germany}}

 ${}^c\!\!$
 {\em {IFAE and Instituci\`o Catalana de Recerca i Estudis Avan\c{c}ats (ICREA) \\ Universitat Aut{\`o}noma de Barcelona
 08193 Bellaterra, Barcelona (Spain)}}

 ${}^d\!\!$
 {\em {HEP Division, Argonne National Laboratory, Argonne, IL 60439}}

 ${}^e\!\!$
 {\em {EFI, KICP and Physics Deparment, Univ. of Chicago, Chicago, IL 60637}}

\end{center}

\vspace{0.8cm}

\centerline{\sc Abstract}
\vspace{2 mm}
\begin{quote}\small
  Electroweak baryogenesis is an attractive scenario for the
  generation of the baryon asymmetry of the universe as its
  realization depends on the presence at the weak scale of new
  particles which may be searched for at high energy colliders. In the
  MSSM it may only be realized in the presence of light stops, and
  with moderate or small mixing between the left- and right-handed
  components. Consistency with the observed Higgs mass around 125 GeV
  demands the heavier stop mass to be much larger than the weak scale.
  Moreover the lighter stop leads to an increase of the gluon-gluon
  fusion Higgs production cross section which seems to be in
  contradiction with indications from current LHC data.  We show that
  this tension may be considerably relaxed in the presence of a light
  neutralino with a mass lower than about 60~GeV, satisfying all
  present experimental constraints.  In such a case the Higgs may have
  a significant invisible decay width and the stop decays through a
  three or four body decay channel, including a bottom quark and the
  lightest neutralino in the final state. All these properties make
  this scenario testable at a high luminosity LHC.
\end{quote}

\vfill

\newpage

\section{\sc Introduction}
\label{introduction}

The origin of the baryon asymmetry is one of the most important open
questions in particle physics and cosmology.  The generation of baryon
asymmetry requires $CP$ and baryon number violation, as well as
non-equilibrium processes~\cite{Sakharov:1967dj}. In the Standard
Model (SM) $CP$ violation is present in the CKM fermion mixing, baryon
number violation is associated with non-perturbative sphaleron
processes and departure from equilibrium may occur at the electroweak
phase transition at finite temperature, below which the Higgs acquires
a vacuum expectation value (VEV).  For this to happen a strongly first
order phase transition should take place in which the vacuum
expectation value of the Higgs $v(T_n)$ at the nucleation temperature
$T_n$ fulfills the condition $v(T_n)/T_n \gtrsim 1$~\cite{reviews}.
However for Higgs masses in the allowed range the electroweak phase
transition in the SM is a cross-over, and therefore the mechanism of
Electroweak Baryogenesis (EWBG) is not realized~\cite{nonpert}.
Moreover even if the phase transition were strong enough, the SM
CP-violating sources are too weak to lead to the observed baryon
asymmetry~\cite{CPSM}.

The realization of the EWBG scenario demands new physics at the weak
scale. Since new physics at this scale is also required for a natural
realization of the Higgs mechanism it is natural to concentrate on
beyond the SM scenarios that fulfill this property. In particular the
Minimal Supersymmetric extension of the Standard Model (MSSM) is a
well motivated one, not only based on symmetry arguments but also
leading to the cancellation of quadratic divergences of the Higgs mass
parameter, to the unification of couplings at high energies, and to a
natural Dark Matter (DM) candidate. Moreover it has been
shown~\cite{CQW,Delepine,CK,FL,JoseR,JRB,Carena:1997gx,Carena:1997ki,CJK,Iiro2,Toni2,Worah,Schmidt,Cline:1998hy,Cline:2000kb,Carena:2000id,Konstandin:2005cd,Cirigliano:2006dg,Carena:2008vj}
that in the presence of a light top squark (stop), with mass lower
than about 120~GeV, and Higgs masses below about 127~GeV (see
Ref.~\cite{Carena:2008vj}) the phase transition can be sufficiently
strong as to allow the realization of this scenario.  New CP-violating
sources may be achieved associated with the light charginos and
neutralinos mass parameters. Therefore all these properties make the
MSSM electroweak baryogenesis scenario testable at current
experiments, in particular in view of the recent observation of
  a Higgs-like resonance with mass close to 125~GeV at both LHC
  experiments~\cite{Gianotti:gia12, Incandela:inc12,
    ATLAS:2012ad,Chatrchyan:2012tw,Atlasnote:-2012-091,CMSnote:12-015,ATLASWW}.

  The Higgs boson mass in the MSSM depends on loop effects, mainly
  associated with the stops.  It depends logarithmically on the stop
  masses, and quadratically and quartically on the mixing parameter in
  the stop sector. For a light stop as the one required for
  baryogenesis, a Higgs boson mass $m_h$ above 115~GeV, consistent
  with the LEP bounds, may only be obtained for a large mixing
  parameter or for a large value of the heaviest
  stop~\cite{Carena:2008rt}.  The large mixing parameter necessary to
  raise the Higgs mass suppresses the coupling of the lightest stop to
  the Higgs and prevents it to have a large effect on the phase
  transition.  Therefore, for EWBG to be realized, the heaviest stop
  mass must be much larger than 1~TeV and the stop mixing parameter
  must be moderate, with values lower than a half of the heaviest stop
  mass~\cite{Carena:2008rt}.

A light stop with small mixing and relevant coupling to the Higgs
tends to enhance by tens of percent, or even factors of a few, the
SM-like Higgs gluon fusion production rate, and somewhat reduces the
decay width into photons, with respect to the
SM~\cite{Djouadi:2005gj,Menon:2009mz}.  Therefore in the decoupling
limit, for large values of the CP-odd Higgs mass, the enhancement of
$\sigma(gg\to h\to\gamma\gamma)$ is somewhat smaller than the
associated enhancement of $\sigma(gg\to h\to WW,\, ZZ)$.  As it has
recently been pointed out by two different
groups~\cite{Cohen:2012zz,Curtin:2012aa} such properties are in
tension with the current Higgs search data at
LHC~\cite{Gianotti:gia12, Incandela:inc12,
  ATLAS:2012ad,Chatrchyan:2012tw,Atlasnote:-2012-091,CMSnote:12-015,ATLASWW}.

In this article we reanalyze the Higgs mass constraints and we will
show that the tension with data may be significantly reduced in the
presence of a light neutralino, with a mass lower than 60 GeV. We
shall show that such scenario is consistent with all present
experimental bounds, in particular with the stop and $Z$-boson decay
constraints.  In this case the Higgs may have a significant invisible
decay width which can compensate the otherwise enhanced $WW$- and
$ZZ$-production rates. In the 125~GeV Higgs-mass region the latter
rates may be close to the ones associated with a SM Higgs, and
therefore consistent with current experimental bounds.

This paper is organized as follows. In section~\ref{LSS} we review the
Light Stop Scenario (LSS) and the conditions for EWBG. In
section~\ref{higgs} we review the constraints from stop and Higgs
searches and study the phenomenological consequences of a light
neutralino.  In section~\ref{correlationsneut} we evaluate the Higgs
production cross sections and decay rates, normalized to the SM ones,
as a function of the neutralino mass for the case of light neutralinos
and supersymmetric parameters consistent with the EWBG conditions
obtained in section~\ref{LSS}. In section~\ref{DM} we analyze possible
Dark Matter candidates in our scenario. Finally we devote
section~\ref{conclusions} to our conclusions.

\section{\sc The Light Stop Scenario and the Electroweak Phase Transition}
\label{LSS}

The realization of the EWBG scenario in the MSSM demands the lighter
stop $(\tilde t)$ to be mainly right-handed and with masses of the
order of 100~GeV.  Since in the MSSM the value of the SM-like Higgs
mass is determined through radiative corrections by the stop masses
$(m_{\tilde t},m_Q)$ and mixing parameter $X_t=A_t-\mu/\tan\beta$,
such a light stop tends to imply the presence of a light Higgs boson
unless the heavy stop mass $m_Q$ is very large. In order to obtain
Higgs boson masses above the LEP limit the heaviest stop mass must be
much larger than 1~TeV.  This implies that a simple one-loop analysis
will not lead to reliable results since it will in general be affected
by large logarithmic functions of ratios of the heavy stop scale to
the weak scale. Such large logarithmic corrections may be efficiently
resummed by means of a Renormalization Group (RG) improvement. In
Ref.~\cite{Carena:2008rt} the technical framework for the treatment of
the light stop scenario, in the presence of a very heavy stop, was
defined by using an effective theory approach and it was subsequently
applied to the EWBG scenario in Ref.~\cite{Carena:2008vj}.  For
completeness, and in order to define a few representative updated
points, we present the results of such an analysis here.

\begin{figure}[htb]
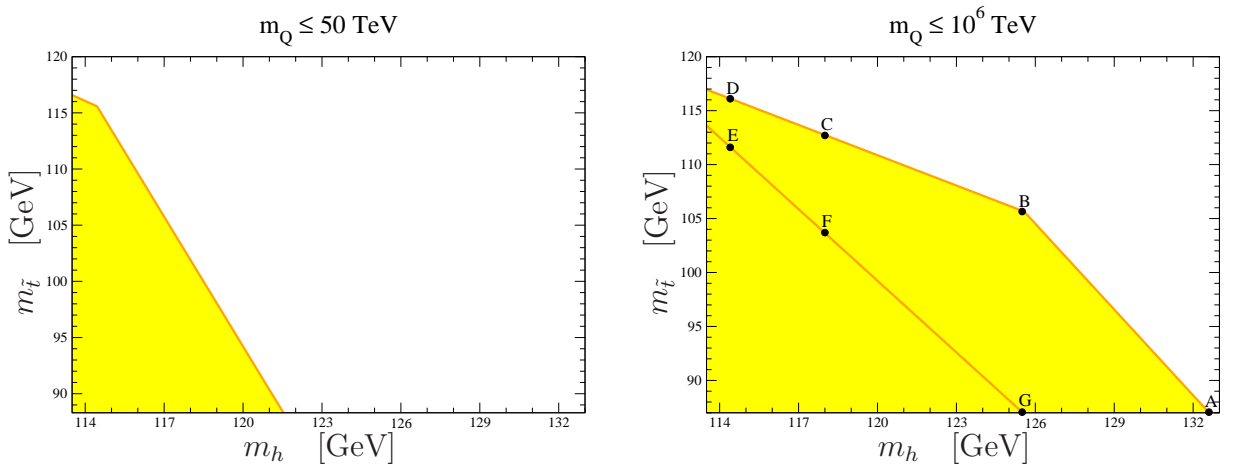

\begin{center}
\psfrag{mh}[c]{$m_h$\quad[GeV]}
\psfrag{mt}[c]{$m_{\tilde t}$\quad[GeV]}
\psfrag{x1}[b]{}
\psfrag{x2}[b]{}
\includegraphics[width=0.47\textwidth]{win50resumGL700.eps}\qquad
\includegraphics[width=0.47\textwidth]{win1e6resumGL700.eps}
\caption{The  window with $\langle\phi(T_n)\rangle/T_n\gtrsim 1$ 
  for a gluino mass $M_3=700$ GeV, $m_Q \le 50\,$TeV (left panel) and
  $m_Q\le10^6$\,TeV (right panel).}
\label{figure-1}
\end{center}
\end{figure}

In order to properly analyze the issue of EWBG we have complemented
the zero temperature results with the two-loop finite temperature
effective potential~\cite{Carena:1997ki}. Light stops may be
associated with the presence of additional minima in the stop--Higgs
$V(\tilde t,h)$ potential, and therefore the question of vacuum
stability is relevant and should be considered by a simultaneous
analysis of the stop and Higgs scalar potentials. All points shown in
Fig.~\ref{figure-1} fulfill the vacuum stability
requirement~\footnote{There is an apparent loss of perturbativity in
  the thermal corrections to the $\tilde t$ potential associated with
  the longitudinal modes of the gluon. In our work we considered that,
  due to their large temperature dependent masses, the terms
  proportional to the third power of their thermal masses in the high
  temperature expansion are efficiently screened and do not lead to
  any relevant contribution to the $\tilde t$ potential.}.

For values of the heavy stop mass $m_Q$ below a few tens of TeV, the
maximal Higgs mass that can be achieved consistent with a strong first
order phase transition is about 122~GeV. The main reason is that
larger values of the Higgs boson mass would demand large values of the
mixing parameter $X_t$, for which the effective coupling
$g_{hh\tilde{t}\tilde{t}}$ of the lightest stop to the Higgs is
suppressed, turning the electroweak phase transition too weak. In the
effective theory the coupling $g_{hh\tilde{t}\tilde{t}}$ is given by
\begin{table}[hbt]
\begin{center}
\begin{tabular}{||c|c|c|c|c|c|c|c||}\hline
Point&A&B&C&D&E&F&G\\
\hline
$|A_t/m_Q|$&0.5&0&0&0&0.3&0.4&0.7\\
\hline
$\tan\beta$&15&15&2.0&1.5&1.0&1.0&1.0\\
\hline
\end{tabular}
\caption{Values of the fundamental parameters at the scale
  $m_Q=10^6$\,TeV corresponding to the benchmark points shown in the left panel of
  Fig.~\ref{figure-1}.}
\end{center}
\label{tabla}
\end{table} 
\begin{equation}
g_{hh\tilde{t}\tilde{t}} \simeq h_t^2\left( 1 - \frac{X_t^2}{m_Q^2}
\right)\left(1+\Delta_g\right)
\label{effcoupl}
\end{equation}
where $\Delta_g$ contains one-loop threshold and radiative corrections
(see Ref.~\cite{Carena:2008rt} and Fig.~1 of
Ref.~\cite{Menon:2009mz}).  Such Higgs mass values, below 122 GeV,
would not lead to an explanation of the Higgs signal observed at the
LHC~\cite{Gianotti:gia12, Incandela:inc12,
  ATLAS:2012ad,Chatrchyan:2012tw,Atlasnote:-2012-091,CMSnote:12-015,ATLASWW}.

For larger values of the heaviest stop mass the logarithmic
corrections to the Higgs mass increase and larger values of the Higgs
mass may be obtained for the same value of $X_t/m_Q$, preserving the
strength of the phase transition.  In this paper we shall focus on
benchmark points where $m_Q=10^6\,$TeV.  This is represented in the
right panel of Fig.~\ref{figure-1}, where it is shown that values of
the Higgs mass as large as 132~GeV may be obtained for this value of
$m_Q$ and (relatively large values of) $\tan\beta \simeq 15$,
corresponding to point A.  However any given point inside the EWBG
region calculated at $m_Q=10^6\,$TeV and moderate $\tan\beta$ can also
be conveniently obtained by decreasing $m_Q$ and increasing
$\tan\beta$. Even for $\tan\beta \simeq 1$ values of the Higgs mass
about 125 GeV may be obtained for $m_Q = 10^6$~TeV, as it is
represented by point G in Fig.~\ref{figure-1}.  The largest values of
the Higgs mass are obtained for the largest possible values of the
Higgs mixing parameter, which in turn leads to the smallest values of
the lightest stop mass consistent with a strong electroweak phase
transition.  Points A and B have $\tan\beta \simeq 15$ while the rest
of the points have smaller values of $\tan\beta$ as shown in Tab.~1,
which defines the values of the fundamental parameters for the
benchmark points used in this work~\footnote{Notice that the
  parameters $A_t\simeq X_t$ as $\mu=\mathcal{O}(100\,{\rm GeV})\ll
  m_Q$ in the LSS.}.
Finally let us stress that, although in this paper we concentrate on
the MSSM case, the value of $m_Q$ can be considerably lowered in some
non-minimal UV completions of the LSS~\cite{Delgado:2012rk}.

\section{\sc Light Neutralinos and the EWBG Scenario}
\label{higgs}

In this section, we shall study the effects of light neutralinos on the 
$Z$ and Higgs invisible width, as well as on the stop phenomenology
within the EWBG scenario.
As it was discussed in section~\ref{introduction}, a light stop with
relevant couplings to the Higgs (leading to a modification of the
phase transition strength) 
has relevant implications for Higgs phenomenology: it induces an
effective enhancement of about a factor two of the diphoton rate
associated with Higgs
production~\cite{Djouadi:2005gj,Menon:2009mz}. Such an enhancement is
not in conflict with data, but even larger enhancements of the $WW$
and $ZZ$ rates are induced, that seem to be in tension with the LHC
data for $m_h\simeq 125$\,GeV~\cite{Cohen:2012zz,Curtin:2012aa}. The
aim of this paper is to extend the analyses of
Refs.~\cite{Cohen:2012zz,Curtin:2012aa} to the case of light
neutralinos with masses below 60\,GeV. This will open the Higgs
(invisible) decay channel into neutralinos and then reduce the
enhancement of the gauge boson rates, thereby alleviating the tension
with LHC data.  As it was already done in Ref.~\cite{Cohen:2012zz} the
study will be performed taking into account the correlation of the
coupling $g_{hh\tilde{t}\tilde{t}}$ with the stop mass $m_{\tilde{t}}$
as well as the requirement of a strong first order phase
transition. On the other hand, as in Ref.~\cite{Curtin:2012aa}, we
will simplify the analysis by considering the values of Higgs
interactions with higgsinos and gauginos to be supersymmetric-like and
not the proper ones of the effective theory~\footnote{Supersymmetry
  implies equality of couplings of several interactions, which are
  modified when the heavy MSSM scalars are decoupled. Estimating the
  numerical value of each effective coupling is
  cumbersome~\cite{Carena:2008rt} and since the couplings involved are
  weak couplings one expects the departures from their precise
  supersymmetric values to lead to subleading effects.}.

\subsection{Collider Constraints on Light Stops}

There have recently been many theoretical analyses on stop mass
limits~\cite{lhcsusy2}. However as most of them do not fit the precise
structure and freedom of the LSS we will consider in this section only
experimental results.

Apart from their effect on Higgs physics, light neutralinos  qualitatively modify the
stop phenomenology, which also depends to a large extent on the mass
of the lightest chargino. Charginos may be heavier or lighter than the
lightest stop:
\begin{itemize}
\item
If the lightest chargino is lighter than the stop, the latter is expected to
decay in a two body decay channel $\chi_1^+b$. The chargino will then
subsequently decay into a $W$  (on-shell or off-shell) and the lightest
neutralino or, in the presence of light third generation sleptons, into a
$\tau$, a neutrino and the lightest neutralino.  The first possibility is
strongly constrained experimentally: a lower bound on the stop mass of about 150~GeV
was obtained under the above assumptions~\cite{CDFstop,ATLASstop, ATLASstop2}. 
\item
If the charginos are heavier than the lightest stop, then stops can
decay through a two body decay channel,
\begin{equation}
\tilde{t} \rightarrow c ~\chi^0_1~,
\label{twobody}
\end{equation}
or through a three (four) body decay channel provided that $m_{\tilde
  t}>m_W+m_b+m_{\chi_1^0}$ ($m_{\tilde
  t}<m_W+m_b+m_{\chi_1^0}$),
\begin{equation}
\tilde{t} \rightarrow b W^+  \chi^0_1\quad (\tilde t\to  b\chi_1^0\bar f f')~,
\label{bWchi}
\end{equation}
where the four-body channel appears through the exchange of a virtual
$\chi^+_1$ or top quark.  The three body decay is in general the
dominant stop decay mode if kinematically allowed in this region of
parameters. In addition, other decays may be present in the case of
light sfermions, that may contribute to the decay amplitude in
channels which do not involve the charged gauge bosons. An example, as
mentioned above, would be a light $\tilde\tau$, or a light
$\tilde{\nu}_{\tau}$ (both heavier than $\chi^0_1$ not to make the
neutralino unstable), which may lead to final states including $\tau$
leptons and neutrinos, apart from a $b$-quark and a
neutralino~\footnote{In our analysis we impose $\tan\beta\le15$ from
  conservative EDM and baryon asymmetry density
  constraints~\cite{Carena:2008rt}. For $\tan\beta\le15$ light staus
  do not essentially modify the Higgs phenomenology presented in this
  paper. However in the presence of cancellations in the EDMs
  chargino-neutralino contributions larger values of $\tan\beta$
  ($\tan\beta\sim60$) would be allowed and consequently light staus
  would enhance the diphoton Higgs decay
  rate~\cite{Carena:2011aa,Carena:2012gp} as well as the baryon
  asymmetry production~\cite{Kozaczuk:2012xv}. We leave this
  possibility for future studies.}.  We have checked that for
off-shell charginos and $m_{\tilde{t}} < m_W + m_b + m_{\chi^0_1}$,
and values of $M_2$ and $\mu$ of order of the weak scale, the decays
$\tilde t \to b \tau^+ \tilde{\nu}_{\tau}$ or $\tilde t \to b
\nu_{\tau} \tilde{\tau}^+$ become the dominant ones whenever one (or
both) of these three body decay channels is kinematically allowed.
\end{itemize}

Searches at the LEP and Tevatron experiments have put very strong
bounds on the lightest stop and neutralino masses using the two body
decay channel \eqref{twobody}. In fact by assuming BR$(\tilde{t}
\rightarrow c ~\chi^0_1)= 1$ LEP imposes $m_{\tilde t}\gtrsim 95$\,GeV
and Tevatron requires $m_{\tilde t}-m_{\chi^0_1}\lesssim 35$
GeV~\cite{Calfayan:2009zz}. Therefore if there are no stop decay
channels competing with the decay $\tilde{t} \rightarrow c ~\chi^0_1$
these experimental bounds imply $m_{\chi^0_1}\gtrsim 60$\,GeV, which
closes the Higgs decay into neutralinos. On the other hand when
BR$(\tilde{t} \rightarrow c ~\chi^0_1)<1$, the stop mass lower bound
becomes weaker. For that reason, and to realize the EWBG scenario (87
GeV $\lesssim m_{\tilde t}\lesssim 120$ GeV), additional light stop
decay channels are required to permit sizable BR$(h \rightarrow
\chi^0_1 \chi^0_1)$~\footnote{See for instance
  Ref.~\cite{Heister:2002hp} for constraints on light stops and
  neutralinos in some scenarios where different channels compete.}.

Interestingly enough, assuming no tree-level flavor violating
couplings the dominant loop-induced contributions to the two body
decay channel, Eq.~(\ref{twobody}), tend to be suppressed in the LSS
discussed in this paper. Therefore, one can consider the possibility
of a four body decay channel as the dominant one when neutralinos are
heavy enough to kinematically forbid the three body decay channel.  To
quantify the previous statement, we will now consider the particular
case of the LSS with a light right-handed stop $\tilde t$, small
mixing in the stop sector $X_t\ll m_Q$, as preferred by the EWBG
mechanism, and light charginos and neutralinos, while the rest of
squarks and heavy Higgses are very heavy with a common mass $m_Q$.
Considering only the contributions enhanced by large logarithmic
factors depending on the ratio of the supersymmetry breaking scale to
the weak scale, the partial width of the decay $\tilde t\to c\chi_1^0$
is given by~\cite{Hikasa:1987db}
\be
\Gamma(\tilde t\to c\chi_1^0)=\frac{\alpha}{4}m_{\tilde t}\left(1-\frac{m_{\chi_1^0}^2}{m_{\tilde t}^2}\right)^2 |f_L \epsilon|^2
\ee
where
\be
f_L=\sqrt{2}\left[ \frac{2}{3}(c_W N_{11}+s_W N_{12})+\left(\frac{1}{2}-\frac{2}{3}s_W^2\right)\left(\frac{N_{12}}{s_w}- \frac{N_{11}}{c_w}\right) \right]
\ee
with $N_{11}$ ($N_{12}$) the  Bino (Wino) component of the lightest neutralino, respectively, and
\be
\epsilon=\frac{\alpha}{4\pi s_W^2}\frac{V_{tb}^*V_{cb} m_b^2}{2 m_W^2 \cos^2\beta}\left[\frac{m_tA_b}{m_Q^2}-\left(3+\frac{A_b^2}{m_Q^2}\right)\frac{m_tX_t}{m_Q^2}\right]
\log\frac{\Lambda^2_S}{m_W^2}
\label{epsilon}
\ee
is a radiatively induced mixing between the light stop and the
left-handed charm squark, with $\Lambda_S$ the messenger scale where
supersymmetry is transmitted to the observable sector.  Notice that
consistently with the assumption of small mixing in the stop sector we
have expanded in Eq.~(\ref{epsilon}) the mixing angle to first order
in the expansion parameter $m_t A_{t,b}/m_Q^2$.  Even for sizable
values of the mixing parameters in the sbottom and/or stop sectors,
$A_t, A_b \simeq {\cal{O}}(m_Q)$, the mixing $\epsilon$ has the extra
suppression $m_t/m_Q$, and hence the partial width $\Gamma(\tilde t\to
c\chi_1^0)$ is very suppressed at the considered order.  The remaining
loop contributions are not enhanced by large logarithmic factors and
therefore in the LSS the four body decay channel can efficiently
compete with the two body decay channel and can become the dominant
one when the three body decay channel is kinematically forbidden.

The three and four body decay final states, Eq.~\eqref{bWchi}, are
similar to those with light charginos previously discussed, but now
the decays proceed through the off-shell chargino production. 
It would therefore be interesting to extend the analysis of
Ref.~\cite{ATLASstop} to the case where the chargino generated by the
stop decay are not on-shell.  The presence of light third generation
sleptons can affect the final state of the stop decay and a careful
analysis of the experimental constraints must be performed considering
the stop decay channel $\tilde{t} \to b \tau^+ \nu_{\tau}
\chi_1^0$. Moreover constraining the stop mass would require different
strategies if the lightest neutralino were unstable, as in the
presence of $R$-parity violation.

Light stops and light neutralinos can also affect the top quark
phenomenology, since the decay channel
\be
t\to \tilde t \chi_1^0
\ee
opens up. This decay channel leads to the decay width~\cite{Hosch:1997vf}  
\be
\Gamma(t\to \tilde t \chi_1^0)\simeq \frac{1}{16\pi}\left(\frac{2 e}{3 c_W}\right)^2m_t\left(1-\frac{m_{\tilde t}^2}{m_t^2}\right)^2
\ee
where for simplicity we have omitted the neutralino mass. In this way
the decay width for the channel $t\to \tilde t \chi_1^0$ is smaller
than about 150\,MeV for $m_{\tilde t}$ in the range 90-115 GeV which
is significantly smaller than the experimental error on the top width
$\Gamma_t=2.0^{+0.47}_{-0.43}$ GeV~\cite{Shary:2012ni}.

\subsection{Higgs and Z Invisible widths}

As mentioned above, we are interested in studying the effect of light
neutralinos on the Higgs production rates in the EWBG scenario. These
effects are induced by a modification of the Higgs decay width.  For
$2 m_{\chi_1^0} < m_h$, the Higgs decay channel into a pair of
lightest neutralinos $\chi^0_1\chi^0_1$ is open with a tree-level
width
\begin{equation}
\Gamma(h\to\chi^0_1\chi^0_1)=\frac{G_F
  m_W^2}{2\sqrt{2}\pi}\,m_h\left(1-\frac{4m_{\chi^0_1}^2}{m_h^2}\right)^{3/2}g_{h11}^2~,
\label{invisib}
\end{equation}
where the coupling of the Higgs to the lightest neutralino, $g_{h11}$,
depends on the product of the gaugino and Higgsino components of the
lightest neutralino. For large values of the pseudo-scalar mass $m_A$
(i.e.~ $\alpha\simeq\beta-\pi/2$), the coupling $g_{h11}$ is given by
\begin{equation}
g_{h11}=(N_{12}-\tan\theta_W N_{11})(\sin\beta N_{1u}-\cos\beta N_{1d})~,
\end{equation}
where $N_{1u}$ and $N_{1d}$ are the neutralino components along the
Higgsino that couples to up and down right-handed quarks, and $N_{11}$
and $N_{12}$ denote the Bino and Wino components, respectively.  Due
to the LEP chargino mass constraints, a lightest neutralino with mass
below 60~GeV must be predominantly Bino.  Hence, the Higgs decay rate
depends on the Higgsino component and gets larger for smaller values
of $\mu$.

On the other hand the relevant Higgsino component, and thus the decay
width $\Gamma(h\to\chi^0_1\chi^0_1)$, becomes more important for
small values of $\tan\beta$, for which the coupling of the lightest
neutralino to the $Z$ boson
\begin{equation}
g_{Z11} = \frac{1}{2} \left(|N_{1u}|^2 - |N_{1d}|^2\right)
\end{equation}
and thus the invisible $Z$ decay width
\begin{eqnarray}
\Gamma( Z \to \chi^0_1\chi^0_1) & = & \frac{G_F}{\sqrt{2} \; 6 \pi} m_Z^3 \left( 1 - \frac{4 m_{{\chi}^0_1}^2}{m_Z^2}\right)^{3/2} g_{Z11}^2 
\nonumber\\
& \simeq & 0.332  \; {\rm GeV} \; g_{Z11}^2 \left( 1 - \frac{4 m_{\chi^0_1}^2}{m_Z^2} \right)^{3/2}~,
\end{eqnarray}
get suppressed~\footnote{Indeed it vanishes for $\tan\beta = 1$.}. 

\begin{figure}[htb]
\begin{center}
\includegraphics[width=0.48\textwidth]{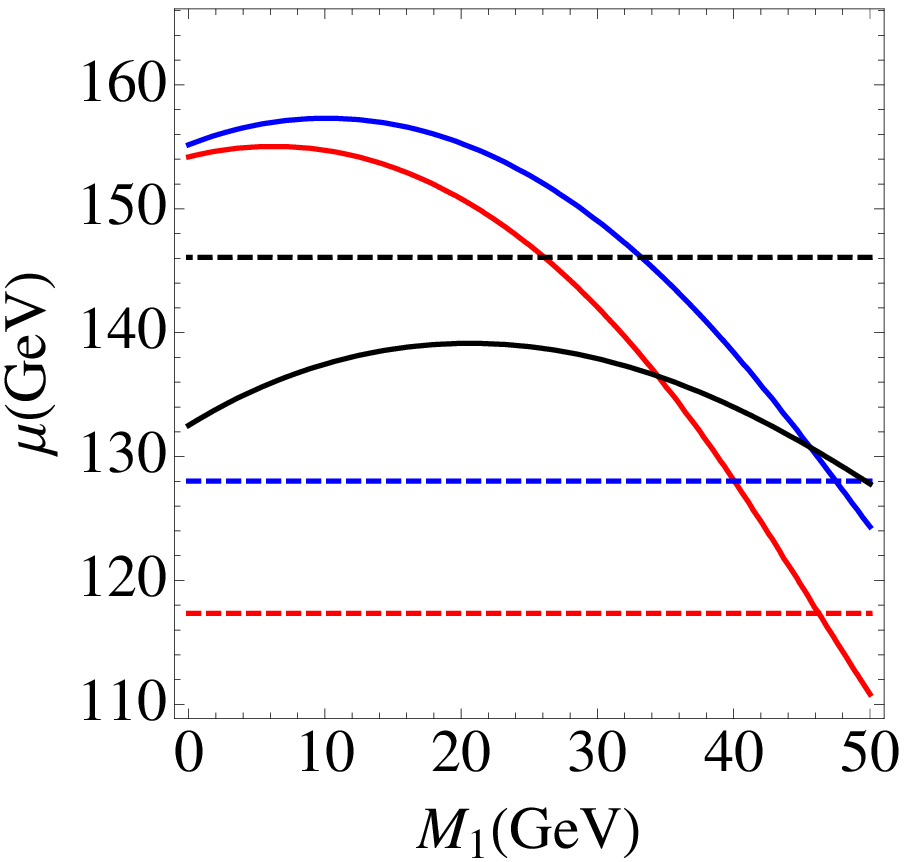}
~~~~\includegraphics[width=0.48\textwidth]{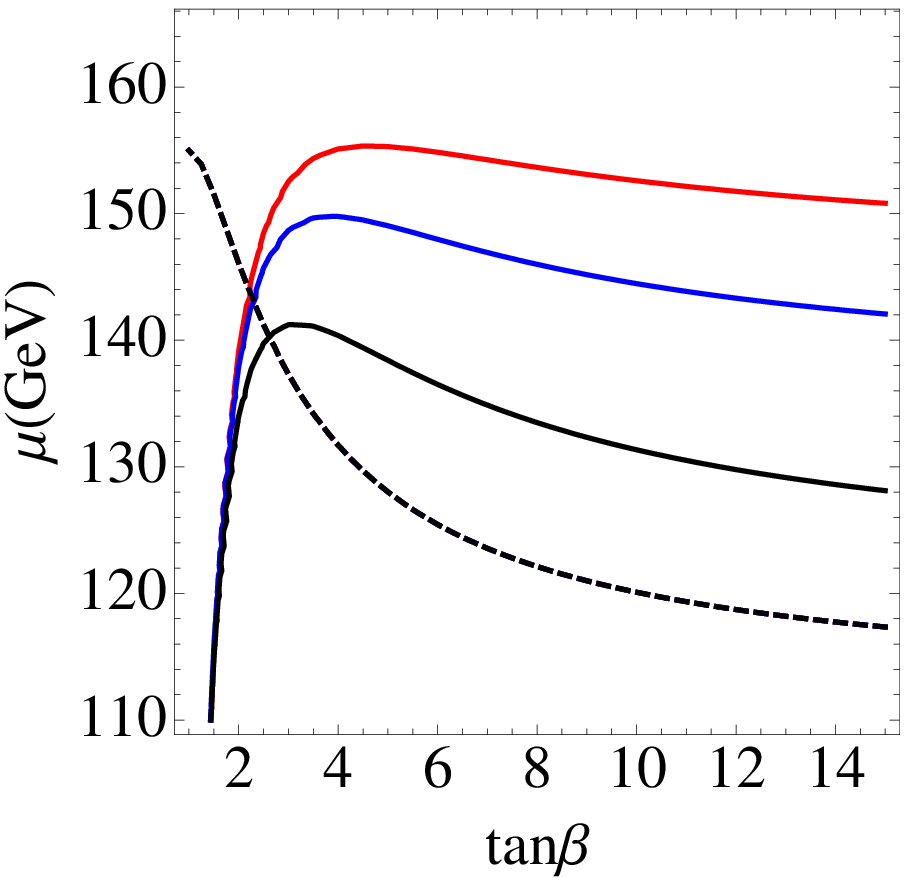}
\end{center}
\caption{Left panel: Allowed region in the $(M_1,\mu)$-plane, for $M_2=200$
  GeV, from the constraint in Eq.~(\ref{invisible}) for $\tan\beta =$ 15 [above
  the intermediate (red) solid line], 5 [above the upper (blue) solid line]
  and 2 [above the lower (black) solid line]. For comparison we also show the
  corresponding allowed regions (above the corresponding dashed lines) for
  $\tan\beta=15$ (lower line) to $\tan\beta=2$ (upper line) from bounds on the
  lightest chargino mass $m_{\chi^\pm_1}>94 $ GeV. Right panel: The same in
  the $(\mu,\tan\beta)$-plane for $M_1=$20 GeV [upper (red) solid], 30 GeV
  [intermediate (blue) solid] and 40 GeV [lower (black) solid]. }
\label{fig:Mubounds}
\end{figure}

Therefore, depending on $\tan\beta$, $\mu$ and $M_2$, the composition of the
lightest neutralino is constrained by the LEP invisible $Z$ width
measurement $\Gamma_{inv}=499.0\pm 1.5 $ MeV~\cite{Nakamura:2010zzi},
which translates into the 95\% CL upper bound
\begin{equation}
\Gamma(Z \to \chi^0_1 \chi^0_1) \lesssim 0.5 \; {\rm MeV}~.
\label{invisible}
\end{equation}
The subsequent lower bound on $\mu$ is shown in
Fig.~\ref{fig:Mubounds} for $M_2=200$ GeV and various values of
$\tan\beta$ (left panel) and $M_1$ (right panel). The solid curves
correspond to the constraint \eqref{invisible} while the dashed lines
correspond to the LEP chargino mass bound $m_{\chi^\pm_1}>94
$~GeV~\cite{Nakamura:2010zzi}. The region below each line in
Fig.~\ref{fig:Mubounds} is excluded at the 95\% CL.\,.
As can be seen from Fig.~\ref{fig:Mubounds}, for small values of $\tan\beta$
the strongest bounds on $\mu$ come from direct searches on charginos,
while for large values of $\tan\beta$ the constraints mainly come from the
  invisible width measurement.  Notice that for $M_2=200$\,GeV, values of $\mu \simeq
200$~GeV are well within the allowed region independently of the value of
$\tan\beta$. The results in Fig.~\ref{fig:Mubounds} are consistent with general
analyses performed in the context of the MSSM~\cite{Dreiner:2009ic}.

On the other hand one can obtain constraints on the invisible Higgs decay
  \eqref{invisib} by looking at signatures of jets plus missing energy. After
  using 1 fb$^{-1}$ of data the
  analysis of this signature by the ATLAS experiment~\cite{ATLASinvHiggs}
   shows that the Higgs production times the
  invisible decay branching ratio must be less than four times the SM Higgs
  production~\cite{TheoryinvHiggs1} which puts no constraint in our
  model. However extrapolating these results to 5 fb$^{-1}$ of data for
  $m_h\simeq 125$\,GeV sets a more stringent constraint on models with
  dominant gluon fusion production~\cite{TheoryinvHiggs2} which, for the LSS,
  would approximately  correspond to
\begin{equation}
  \frac{\sigma(gg \to h)}{\sigma(gg\to h)_{SM}} \times {\rm BR}(h \to {\rm
    inv}) \lesssim 1.9 \;\; {\rm~ at } \; 95\%~{\rm C.L.\;
    }~(m_h\simeq 125\,{\rm GeV})~ .
    \label{hinvbounds}
\end{equation}
In practice this bound would require cross section times the dominant
invisible decay branching ratio enhancement to be smaller than 2.  In most of
the cases analyzed in this article the cross section enhancement is slightly
about a factor 2 and the branching ratio not larger than 0.85 and hence this
constraint tends to be fulfilled (although there are some residual regions in
the parameter space which are not compatible with it).  On the other hand this
bound allows regions that can be already excluded because of other
observables. In particular in the LSS values of BR$(h \to {\rm inv}) \gtrsim
0.7$ lead to Higgs decay rates into SM particles that are in general
significantly smaller than the experimental ones (see
Figs.~\ref{figure-4}--\ref{fig:mufunc})~\footnote{This observation mirrors the
  result of the ATLAS analysis fitting multiple decay channels and leading to
  BR$(h \to {\rm inv})\lesssim 0.84$ at 95\% C.L.~\cite{ATLASinvHiggs2} after
  profiling on gluon fusion and diphoton enhancement factors.  Indeed for any
  model predicting particular values of the gluon and diphoton enhancement
  factors we expect a stronger bound on BR$(h \to {\rm inv})$ than the ATLAS
  one. In particular for the LSS model analyzed in this paper we have checked
  that there is an approximate upper bound on BR$(h \to {\rm inv})\lesssim
  0.7$ as we have mentioned. }.  We shall further comment on this in the next
section.

As a final remark, we stress that the above constraints hold also for
unstable neutralinos provided their lifetime is much larger than the
size of the detector.
  This situation can arise, for example, in the presence of R-parity violation.
  For instance, if the R-parity violation
  affects only the squarks, the stop can gain an extra two-body decay
  channel with a width that, depending on the coupling strength,  may be comparable 
  or larger than the one of the usual stop decay modes, and therefore the bounds discussed
  in section~\ref{LSS} would have to be revised.    
  The neutralino, instead, can have a five-body decay via off-shell top
  and stops, and an off-shell W from the top. The additional number of
  particles in the decay,  together with hypercharge and
  weak couplings, leads to a $\mathcal O(10^6)$ suppression of the neutralino width
  compared with the stop one. The R-parity violation can
  thus modify the stop phenomenology with neutralinos that may remain 
  stable at collider scales, and hence the products of the Higgs decay into 
  neutralinos will remain invisible~\footnote{Even if both stop and neutralino
      would decay promptly, a five body decay of a light neutralino
      necessarily implies soft decay products. These would therefore not appear
      in standard Higgs searches and these decay channels will thus
      practically remain ``invisible".}.

\section{\sc LHC Higgs signatures correlation and dependence on $m_{\chi^0_1}$}
\label{correlationsneut}
As previously discussed, in the EWBG scenario, light stops with
relevant couplings to the Higgs induce modifications to the rates of
gluon fusion Higgs production and Higgs decay into gluon and photon
pairs, which tend to be significant. In particular deviations from the
SM production and decay rates yield some tension with LHC
data~\cite{Gianotti:gia12, Incandela:inc12,
  ATLAS:2012ad,Chatrchyan:2012tw,Atlasnote:-2012-091,CMSnote:12-015}
for $m_h\simeq 125$ GeV~\cite{Cohen:2012zz,Curtin:2012aa}. In
comparison with Ref.~\cite{Curtin:2012aa} our results present
  smaller deviations of the gluon fusion Higgs production rates from
  the SM values. This is due to the fact that we consider the proper
  correlation between the Higgs-stop effective coupling
  $g_{hh\tilde{t}\tilde{t}}$ and the stop mass $m_{\tilde{t}}$ through
  the mixing $A_t$, and that we do not neglect the stop mixing effects
  in the loop-induced production and decay rates. As an example of the
  relevance of such effects, the lightest stops in the EWBG scenario
  are obtained through relatively large $X_t/m_Q$.  In such a case,
  the large enhancement of the gluon production rate obtained by small
  $m_{\tilde{t}}$ is partially suppressed by the reduction of
  $g_{hh\tilde{t}\tilde{t}}$ in Eq.~\eqref{effcoupl}. 
   
\begin{figure}[h]
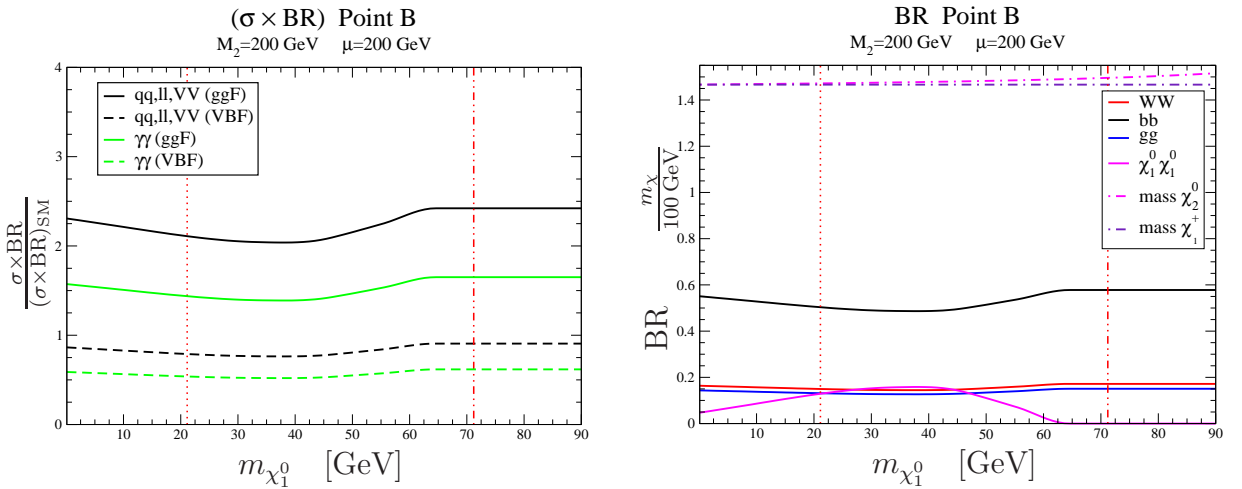

\begin{center}
\psfrag{mh}[c]{$m_h$\quad[GeV]}
\psfrag{mt}[c]{$m_{\tilde t_R}$\quad[GeV]}
\psfrag{x}[c]{$m_{\chi^0_1}$\quad[GeV]}
\psfrag{y2}[c]{$\frac{\sigma\times {\rm
      BR}}{(\sigma\times
    \rm BR)_{\rm SM}}$}
\psfrag{y}[c]{${\rm BR}\qquad~ \frac{m_{\chi}}{100\,{\rm GeV}}$}
\includegraphics[width=0.47\textwidth]{brM200mu200_B.eps}\qquad
\includegraphics[width=0.47\textwidth]{decM200mu200_B.eps}
\caption{$(\sigma\times$BR$)/(\sigma\times$BR$)_{\rm SM}$ of the Higgs
  (left panel) and BR of the Higgs (right panel, channels with
  BR$<0.1$ are omitted) as a function of $m_{\chi^0_1}$ at point B
  of Fig.~\ref{figure-1} for $M_2=\mu=200\,$GeV. The Higgs mass is about 125~GeV. The lightest
  chargino and next-to-lightest neutralino (lower and upper dot-dashed lines in the
  right panel) are heavier than the light stop. The vertical
  dot-dot-dashed line corresponds to the Tevatron lower bound on
  the lightest neutralino assuming BR$(\tilde{t} \rightarrow c
  ~\chi^0_1)=1$. On the left (right) of the vertical dotted line the stop can
  decay as in Eq.~(\ref{bWchi}) with a real (virtual) W boson.}
\label{figure-4}
\end{center}
\end{figure}

\begin{figure}[h!]
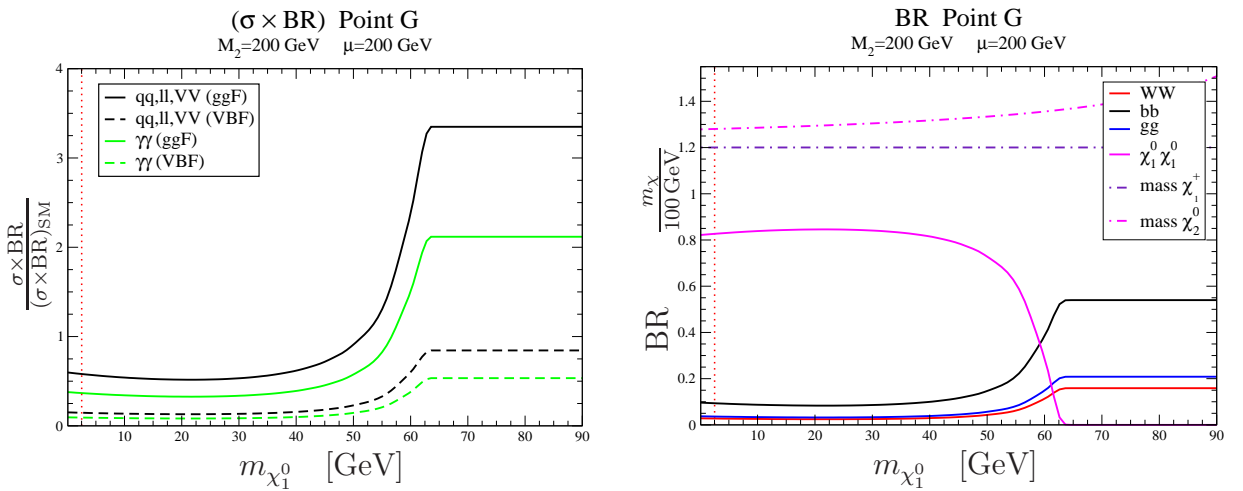

\begin{center}
\psfrag{mh}[c]{$m_h$\quad[GeV]}
\psfrag{mt}[c]{$m_{\tilde t_R}$\quad[GeV]}
\psfrag{x}[c]{$m_{\chi^0_1}$\quad[GeV]}
\psfrag{y2}[c]{$\frac{\sigma\times {\rm
      BR}}{(\sigma\times
    \rm BR)_{\rm SM}}$}
\psfrag{y}[c]{${\rm BR}\qquad~ \frac{m_{\chi}}{100\,{\rm GeV}}$}
\includegraphics[width=0.47\textwidth]{brM200mu200_G.eps}\qquad
\includegraphics[width=0.47\textwidth]{decM200mu200_G.eps}
\caption{The same as Fig.~\ref{figure-4} but for point G and $M_2=\mu=200$ GeV. The Higgs mass is about 125~GeV.}
\label{figure-5}
\end{center}
\end{figure}

\begin{figure}[h!]
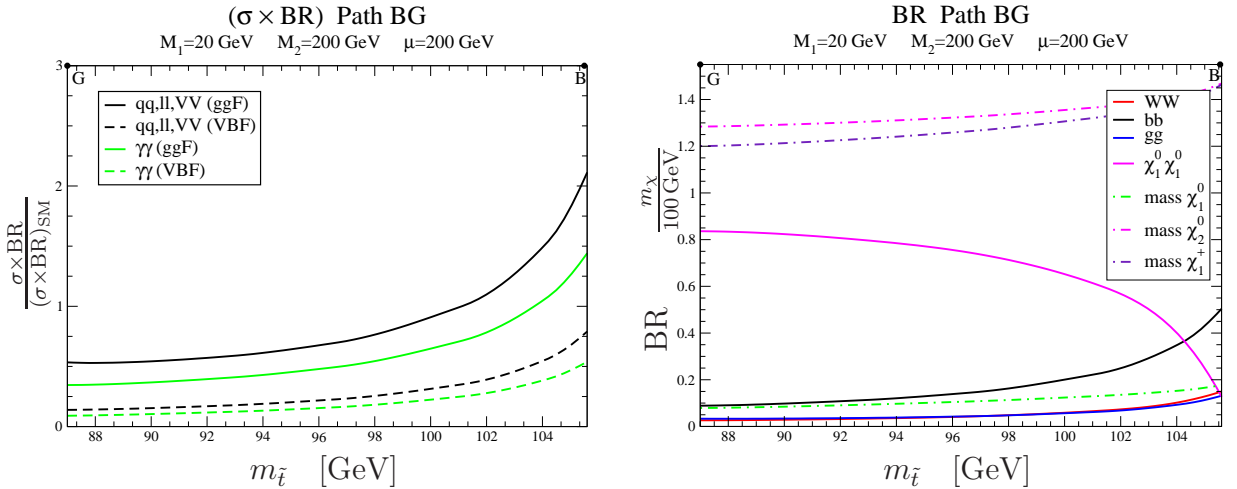

\begin{center}
\psfrag{mh}[c]{$m_h$\quad[GeV]}
\psfrag{mt}[c]{$m_{\tilde t_R}$\quad[GeV]}
\psfrag{x}[c]{$m_{\tilde t}$\quad[GeV]}
\psfrag{y2}[c]{$\frac{\sigma\times {\rm
      BR}}{(\sigma\times
    \rm BR)_{\rm SM}}$}
\psfrag{y}[c]{${\rm BR}\qquad~ \frac{m_{\chi}}{100\,{\rm GeV}}$}
\includegraphics[width=0.47\textwidth]{brM2_200mu200M1_20_mst_BG.eps}\qquad
\includegraphics[width=0.47\textwidth]{decM2_200mu200M1_20_mst_BG.eps}
  \caption{Evolution along the path BG of Fig.~\ref{figure-1} of
    $\sigma\times {\rm BR}/(\sigma\times {\rm BR})_{SM}$ (left panels) and BR (right panels) for $M_2=\mu=200$ GeV and $M_1=20$ GeV.}
\label{figure-10}
\end{center}
\end{figure}

\begin{figure}[h!]
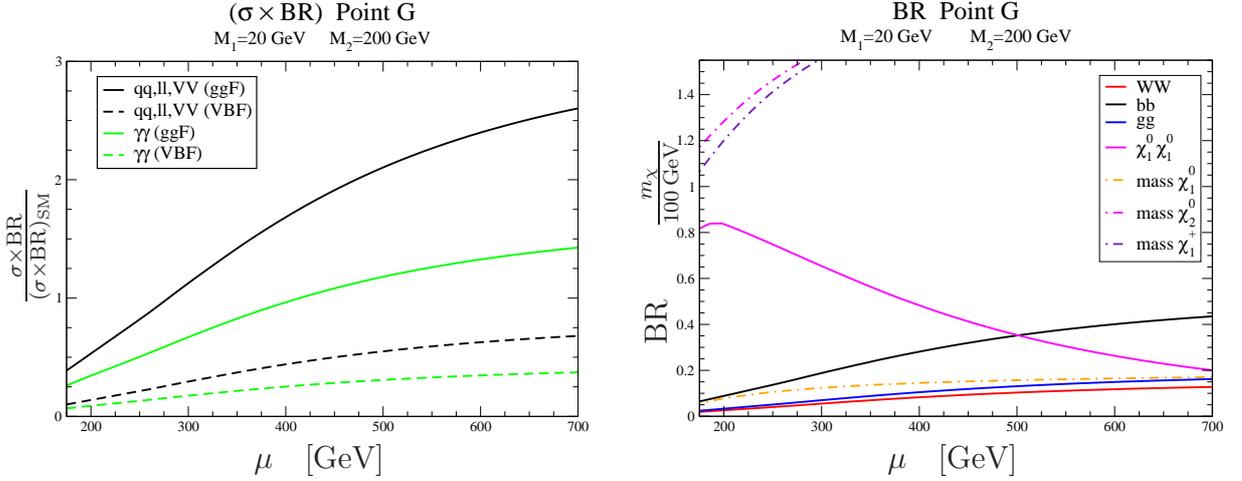

\begin{center}
\psfrag{mh}[c]{$m_h$\quad[GeV]}
\psfrag{mt}[c]{$m_{\tilde t_R}$\quad[GeV]}
\psfrag{x}[c]{$\mu$\quad[GeV]}
\psfrag{y2}[c]{$\frac{\sigma\times {\rm
      BR}}{(\sigma\times
    \rm BR)_{\rm SM}}$}
\psfrag{y}[c]{${\rm BR}\qquad~ \frac{m_{\chi}}{100\,{\rm GeV}}$}
\includegraphics[width=0.47\textwidth]{brM2_200M1_20_G.eps}\qquad
\includegraphics[width=0.47\textwidth]{decM2_200M1_20_G.eps}
\caption{$(\sigma\times$BR$)/(\sigma\times$BR$)_{\rm SM}$ of the
    Higgs (left panel) and BR of the Higgs (right panel, channels with
    BR$<0.1$ are omitted) as a function of $\mu$ at point G of
    Fig.~\ref{figure-1} for $M_1=20\,$GeV and $M_2=200\,$GeV. The
    Higgs mass is about 125~GeV. The lightest chargino [dot-dashed
    (indigo) line in the right panel] is heavier than the stop.}
\label{fig:mufunc}
\end{center}
\end{figure}

In this section we shall
  re-analyze the LSS Higgs phenomenology studied in
  Refs.~\cite{Cohen:2012zz,Curtin:2012aa}
and determine how the Higgs signatures are affected by the
  additional presence of a light neutralino.
In order to study the light neutralino effects we consider values of
the supersymmetric parameters that are consistent with the bounds on
the $Z$ invisible width (\ref{invisible}), as shown in
Fig.~\ref{fig:Mubounds}.  Results for several benchmark points
exhibited in Fig.~\ref{figure-1} are presented in
Figs.~\ref{figure-4}--\ref{fig:mufunc}. On the right panels of these
figures we present the dependence on the neutralino mass
$m_{\chi^0_1}$ of the relevant Higgs decay branching ratios, while in
the left panels we present the ratios of the gluon fusion and weak
boson fusion production cross sections times the branching rates for
SM channels with respect to their corresponding values in the SM. In
the right panels we also plot the next-to-lighest neutralino and
lighest chargino masses, $m_{\chi^0_2}$ and $m_{\chi^+_1}$. We
consider large values of $m_A$ for which the tree-level coupling of
the Higgs to the SM fields is the same as in the SM. Therefore, the
ratio of weak boson fusion production times tree level Higgs decay
rates to SM ones is simply given by the quotient of the branching
ratios.  Instead the ratios of gluon fusion production induced
processes are strongly modified by the presence of light stops.  A few
general comments are here in order.

\begin{itemize}
\item 
The considered cases in this section have charginos always off-shell in
  the decay chain $\tilde t\to b\chi^+_1\to bW^+\chi^0_1$, or similarly
  $\tilde t \to b \tilde{\tau}^+ \nu_{\tau}$ ($\tilde t \to b \tilde \nu_\tau \tau^+$) in the presence of a light
  stau (tau sneutrino). 
\item For $M_2 = 200$\, GeV, values of $\mu \gtrsim 180$\,GeV lead to
  consistency with all experimental constraints and to chargino masses
  such that the two body decay of the stop into an on-shell chargino
  and a bottom quark is forbidden in the whole parameter space under
  study.
\item 
The effect of the neutralino on Higgs physics is much stronger for small $\tan\beta$ due
  to an increase in the Higgsino component of the lightest neutralino. On the
  other hand for the same mass parameters, $m_{\chi^\pm_1}$ and
  $m_{\chi^0_2}$ tend to decrease in value.
\end{itemize}

Fig.~\ref{figure-4} shows the masses $m_{\chi^+_1}$ and
$m_{\chi^0_2}$, the Higgs production rates normalized to the SM ones
and the Higgs decay branching ratios at point $B$, with $\tan\beta =
15$, as a function of the neutralino mass. We have chosen $\mu =
200$~GeV, which widely overcomes the lower bound on this parameter
from the invisible $Z$-width for this value of $\tan\beta$.  For
$m_{\chi^0_1}$ smaller (larger) than 20\,GeV the three body decay
\eqref{bWchi} is allowed (forbidden). The Higgs mass is about 125.5
\,GeV, consistent with the LHC observation.

For $m_h \simeq 125.5$~GeV, both the LHC and Tevatron data are overall
compatible with SM Higgs rates within statistical errors. However,
both ATLAS and CMS see an enhancement in the diphoton
channel~\cite{Gianotti:gia12, Incandela:inc12,
  ATLAS:2012ad,Chatrchyan:2012tw,Atlasnote:-2012-091,CMSnote:12-015},
with the best fit to the diphoton production cross section being
$(1.90 \pm 0.50)$ and $(1.56 \pm 0.43)$ times the SM one, for a Higgs
mass 126.5~GeV and 125~GeV, respectively.  CMS and ATLAS also report
results discriminating between the vector boson fusion and gluon
fusion production channels. CMS shows enhancements of order 2 times
and 1.5 times the SM cross section in the weak boson fusion and gluon
fusion production channels, respectively, but the errors are large and
both channels are only about 1$\sigma$ above the SM predictions.
Similar results are obtained at ATLAS, which shows central values of
order 2 times the SM cross sections in both production channels.  In
the $ZZ$ channel, ATLAS and CMS are in good agreement with SM
predictions~\cite{Gianotti:gia12, Incandela:inc12,
  ATLAS:2012ad,Chatrchyan:2012tw,Atlasnote:-2012-091,CMSnote:12-015},
but with rates about $(1.3 \pm 0.6)$ and $\left(0.7^{+0.5}_
  {-0.4}\right)$ times the SM one, and hence also consistent with
slight suppressions or enhancements of these rates. Similarly, the
best fit to the ATLAS and CMS $WW$ production
rates~\cite{Gianotti:gia12, Incandela:inc12,
  ATLAS:2012ad,Chatrchyan:2012tw,Atlasnote:-2012-091,CMSnote:12-015,ATLASWW}
are about $(1.4 \pm 0.5)$ and $(0.6^{+0.5}_{-0.4})$ times the SM one,
respectively.  CMS also shows a large suppression of $WW$ production
in the vector boson fusion channel, but with a very large error.  CMS
also reports a suppression of $\tau \tau$ production in the vector
boson fusion channel\cite{Incandela:inc12,CMSnote:12-015}. No such
suppression is seen in the gluon fusion channel. Overall, considering
all the production and decay channels explored at the LHC, the best
fit performed at CMS shows a suppression of the vector boson fusion
induced rates with respect to those expected in the SM and gluon
fusion induced rates that are consistent with the SM ones. As we will
show, such overall behavior is consistent with the predictions of the
LSS in the presence of light neutralinos.

As it is highlighted in Fig.~\ref{figure-4}, for
$m_{\chi{^0_1}}\gtrsim 63\,$GeV the Higgs cannot decay into
neutralinos. In such a case the Higgs production via gluon fusion is
enhanced by a factor larger than two. Then the subsequent Higgs decay
into weak bosons, whose rate is unmodified by light stops at leading
order, is enhanced by the same factor of two. This enhancement factor
is instead suppressed by $\sim$ 25\% if the Higgs decays into photons
because of the stop destructive-interference contribution.  
If $m_{\chi{^0_1}}\lesssim 63\,$GeV the Higgs invisible width
increases. However for relatively large values of $\tan\beta$, as
point B, and for $\mu=M_2= 200\,$GeV, the coupling $g_{h11}$ is
suppressed, and opening kinematically the Higgs decay channel into
neutralinos reduces the visible branching ratios by at most 10\%.  
In conclusion, for point B we confirm the result of the previous
analyses~\cite{Cohen:2012zz,Curtin:2012aa}: for heavy neutralinos
there is tension with data, independently of the specific choice of
$\mu$ and $M_2$. Moreover, for smaller values of the neutralino mass
the tension persists, unless one assumes smaller values of $\mu$
and/or $M_2$ than those considered here, $\mu=M_2= 200\,$GeV (see
comments on Fig.~\ref{figure-11} for more details).

In order to see at work the mechanism of invisible Higgs decay into
neutralinos for $\mu\approx M_2\approx 200\,$GeV, we have to consider
small values of $\tan\beta$ for which the coupling $g_{h11}$ is
sizable. In particular this is the case for point G in
Fig.~\ref{figure-1}.  Fig.~\ref{figure-5} shows the corresponding
results for point G, for which $\tan\beta$ is close to 1 and the Higgs
boson mass is still about 125~GeV~\footnote{With respect to point B
  the smaller tree-level Higgs mass due to the decrease of $\tan\beta$
  is compensated by rising the mixing parameter $X_t$ in the radiative
  contributions.}. In this case the Higgsino $\tilde{H}_u$ component
of the lightest neutralino increases and hence the decay width of the
Higgs into the lightest neutralino can be more significant even for
$\mu = M_2=200$~GeV. Indeed in the region in which the Higgs is
allowed to decay into neutralinos its branching ratio tends to be the
dominant one. On the other hand in the large $m_{\chi^0_1}$ region one
obtains enhancement factors in the Higgs production rates mediated via
gluon fusion which are even larger than for point B.  These large
enhancement factors are, however, compensated by the large increase of
the Higgs invisible width for small values of the neutralino
mass. Values of the vector-boson production rate about 60\% (or less)
the SM value are obtained in the gluon fusion induced channels for
small neutralino mass.  The weak boson fusion production signatures
are strongly suppressed with respect to the SM case.  Such values are
in tension with present
data as the predicted diphoton rate is small compared to the current
results at this invariant diphoton mass in both the gluon fusion
channel as well as in the channel with two jets, in which vector boson
fusion production contributes in a significant way. Moreover, the rate
of invisible Higgs production is large, being in tension with the
current bounds on this rate, Eq.~(\ref{hinvbounds}). In conclusion, at
point G, for heavy neutralinos the enhancement of signatures induced
through gluon fusion is in tension with data, as already observed in
the literature. The suppression mechanism via light neutralinos leads
to both gluon fusion and vector boson fusion induced rates that are
too small, and also in tension with data. Values of the neutralino
mass in the range 55-57~GeV lead to the best description of data, and
require extra stop decay channels beyond the $c\chi^0$ and
$W^+b\chi^0$ ones to be consistent with stop physics.  The predicted
rates of the signatures induced through vector boson fusion are about
half of the SM values.  In order to compare with data, however, a
better understanding of the dijet channel rates coming from the gluon
fusion Higgs production, whose rate in the LSS is significantly larger
than in the SM, must be achieved.

As we have seen from Figs.~\ref{figure-4} and \ref{figure-5} the
dilution effect of the Higgs invisible decay, that is mainly governed
by $\tan\beta$ (when the channel is kinematically accesible), vanishes
for point B ($\tan\beta=15$) and it is maximized for point G
($\tan\beta\simeq 1$). Both points are hence in tension with data for
very light neutralinos. Along the path BG this effect varies
continuously and one can find all intermediate cases. We illustrate
the variation of this effect in Fig.~\ref{figure-10} which shows the
Higgs production rates with respect to the SM values (left panel) and
the Higgs branching ratios (right panel) as a function of the stop
mass along the path BG~\footnote{Many paths in the full parameter
  space have the same projection in the plane $(m_h,m_{\tilde t})$ and
  fulfill the condition $v(T_n)/T_n\gtrsim1$. The path we have
  considered is that with $v(T_n)/T_n\simeq1$.}.  In
Fig.~\ref{figure-10}, where $m_{\chi^0_1}\approx 15\,$GeV, better
compatibility with LHC data is reached at larger stop masses
($m_{\tilde t}\approx 104$\,GeV with $\tan\beta\simeq 5$). No
significant variation of these results would be obtained for
$m_{\chi_1^0} \simeq 40$\,GeV, for which the three body $W^+b\chi_1^0$
decay channel would be kinematically forbidden.

For smaller stop masses (larger stop mixing) along the path BG,
$\tan\beta$ decreases and the Higgs branching ratios to the visible
sector are excessively reduced because of the large invisible decay
width.  Nevertheless, this does not exclude the EWBG region at small
stop mass values. Better agreement with data at smaller stop masses
can be achieved by reducing BR($h\rightarrow\chi{^0_1}\chi{^0_1}$) in
several ways: \textbf{i)} By assuming larger $m_{\chi^0_1}$;
\textbf{ii)} By reducing $m_Q$, which allows to increase $\tan\beta$
for a given $m_{\tilde{t}}$; and, \textbf{iii)} By considering larger
$\mu$ and to a lesser extent $M_2$, as it is illustrated in
Fig.~\ref{fig:mufunc} for point G. We can see that, for the values of
$M_1$ and $M_2$ considered in Fig.~\ref{fig:mufunc}, better agreement
with experimental data can be achieved for values of $\mu$ in the
range $300\, \textrm{GeV}\lesssim\mu\lesssim \, 400$ GeV.  The ratio
of the $h \to ZZ$ rate to the $h \to \gamma\gamma$ rate becomes larger
for larger values of $\mu$, due to a suppression of the chargino
effects, which tends to compensate the negative contributions to the
$h \to \gamma \gamma$ amplitude induced by the light stops.
    
In general, we see that once the light neutralino effects are
considered, consistency of the $gg \to h \to ZZ,WW, \gamma\gamma$
results with experimental data~\cite{Gianotti:gia12, Incandela:inc12,
  ATLAS:2012ad,Chatrchyan:2012tw,Atlasnote:-2012-091,CMSnote:12-015}
may be restored.  Higgs invisible decay branching ratios of order
30~--~60~\% lead to the best description of current LHC Higgs
results. These invisible width contributions may be obtained by
adjusting the value of $\mu$, and hence the Higgsino component, in the
case of very light neutralinos, or by taking values of the neutralino
mass close to $m_h/2$, for low values of $\mu$.  For fixed $\mu$ and
$m_{\chi_1^0}$, one can also adjust $\tan\beta$ by varying $m_Q$ and
hence obtain the desirable invisible width for $m_h \simeq 125.5$~GeV.
All vector boson fusion induced channels tend to be suppressed due to
the increase of the Higgs width, and this is consistent with the
overall behavior observed at the CMS experiment.  However, the
experimental value of the rate of the dijet $h \to \gamma \gamma$
channel at the ATLAS and CMS experiments is currently larger than the
SM one at the 1$\sigma$ level.  The main contribution to this channel
at ATLAS and CMS is expected to come from vector boson fusion
production.  Therefore, the vector boson fusion channel $h\to \gamma
\gamma$ imposes the strongest constraint on the realization of the LSS
with light neutralinos within current
experimental data. However, as stressed above, the contribution to the
dijet channel from gluon fusion production may be larger in the LSS
than in the SM, and therefore imparing a naive comparison of the
vector boson fusion predictions with the dijet channel data.

\section{\sc Dark Matter}
\label{DM}
A relevant question regarding the lightest neutralino we are
considering is whether it can be a thermal WIMP.  The low values of
its mass determine that the neutralino co-annihilation with stops and
charginos becomes subdominant, as well as the s-channel annihilation via
Higgs bosons. The only relevant channel then is the one mediated by a
$Z$ gauge boson. However constraints on the invisible $Z$ width
determine that this channel might not always be efficient enough to
accomodate the cosmological thermal DM abundance of the
Universe. 

\begin{figure}[h!]
\begin{center}
\includegraphics[width=0.48\textwidth]{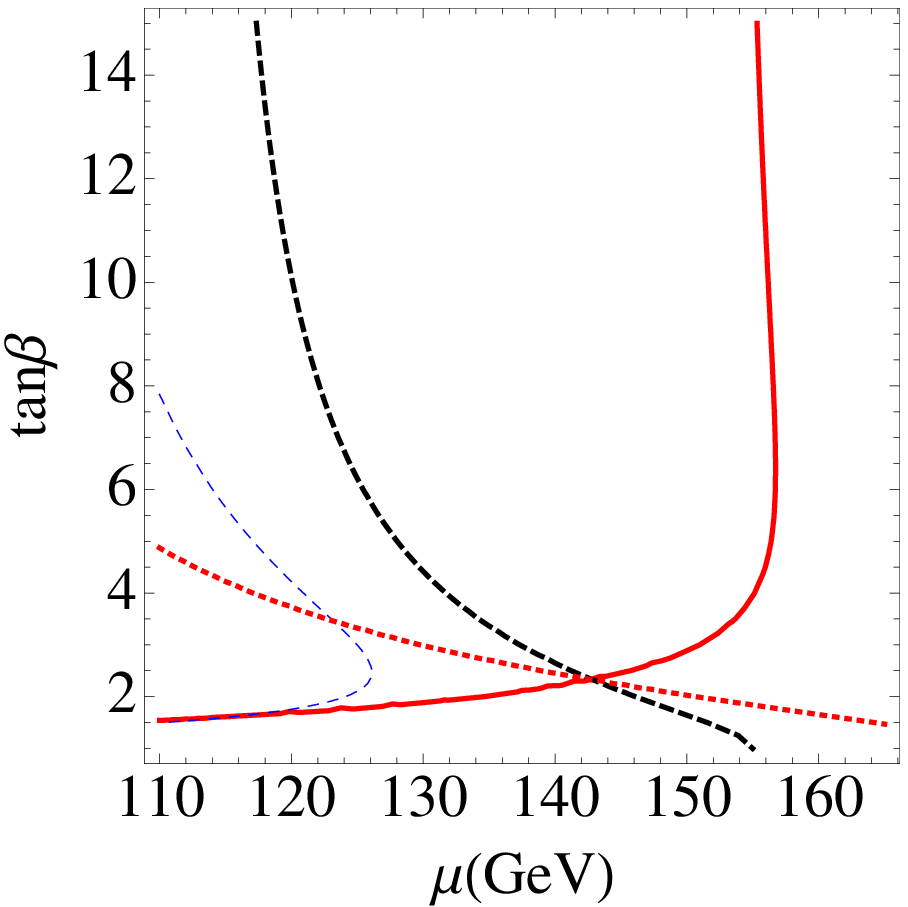}
~~~~\includegraphics[width=0.48\textwidth]{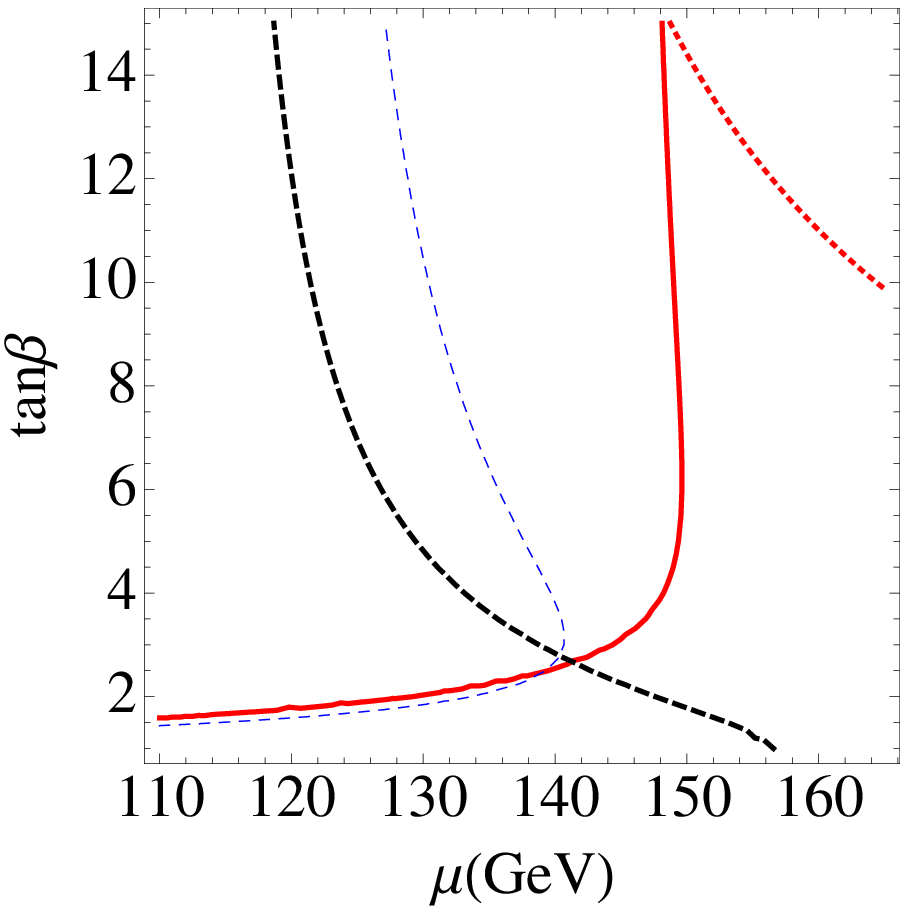}
\caption{Left panel: Contour plots of $g_{Z11}=0.05$ [thick solid
  (red)] for $M_2=200$ GeV and $M_1=55$ GeV, $m_{\chi_1^+}=95$ GeV
  [thick dashed (black): allowed region on its right],
  $m_{\chi_1^0}=35$ GeV [thin dotted (red)] and
  $\Gamma(Z\to\chi_1^0\chi_1^0)$=0.5 MeV [thin dashed (blue): allowed
  region on its right]. Right panel: The same as in the left panel but
  with $M_1=40.6$ GeV. }
\label{fig-dm}
\end{center}
\end{figure}

An example that is compatible with the observed DM
arises when, for instance, $m_{\chi_1^0}=35-40$ GeV for $g_{Z11}
\simeq 0.05$~\cite{MMW}~\footnote{Notice that we can consistently
  produce the observed thermal DM density for larger (smaller) values
  of $m_{\chi_1^0}$ if $g_{Z11} \lesssim 0.05$ ($g_{Z11} \gtrsim
  0.05$) provided that condition (\ref{invisible}) is
  satisfied.}. These two conditions can be easily reproduced in the
LSS, as it is shown in Fig.~\ref{fig-dm}. In the left panel of
Fig.~\ref{fig-dm} we plot the contour line $g_{Z11}= 0.05$ in the
$(\mu,\tan\beta)$-plane for $M_2=200$\,GeV and $M_1=55$\,GeV. Taking
the constraint $m_{\chi_1^+}\gtrsim 95$\,GeV~\cite{Nakamura:2010zzi},
we exclude the region on the left of the thick dashed (red) curve. As
the plot shows this constraint is stronger than condition
(\ref{invisible}) [excluding the region on the left of the thin dashed (blue) curve] for
the considered parameter choice.  Notice that the contour lines
$m_{\chi_1^0}=35$ GeV [thin dotted (red) curve],
$m_{\chi_1^+}=95$\,GeV and $g_{Z11}= 0.05$ cross at the same point,
which means that for $M_2=200$\,GeV the parameter $M_1$ needs to be
smaller than 55\,GeV in order to fulfill the above conditions on
$g_{Z11}$, $m_{\chi_1^+}$ and $\Gamma(Z\to\chi_1^0\chi_1^0)$. On the
other hand by assuming $\tan\beta\lesssim 15$ as imposed by
EWBG~\cite{Carena:2008vj} one finds that the conditions
$M_2=200$\,GeV, $m_{\chi_1^0}=35$ and $g_{Z11} \simeq 0.05$ cannot be
fulfilled with $M_1$ smaller than $40.6\,$GeV (case plotted in the
right panel of Fig.~\ref{fig-dm}). Moreover for such a value of $M_1$
the constraints on $\Gamma(Z\to\chi_1^0\chi_1^0)$ and $m_{\chi_1^+}$
are satisfied.

\begin{figure}[h!]
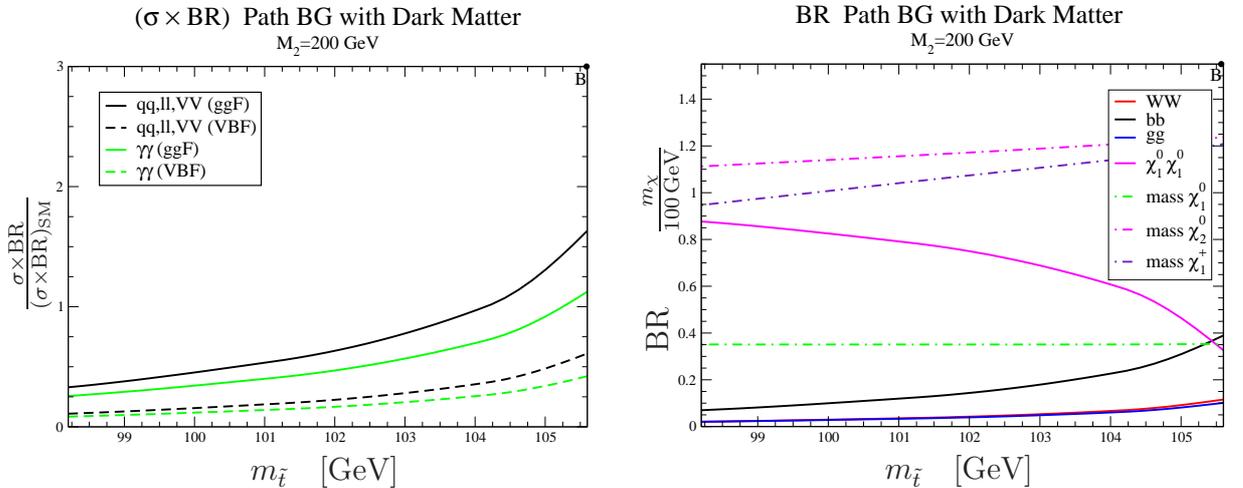

\begin{center}
\psfrag{mh}[c]{$m_h$\quad[GeV]}
\psfrag{mt}[c]{$m_{\tilde t_R}$\quad[GeV]}
\psfrag{x}[c]{$m_{\tilde t}$\quad[GeV]}
\psfrag{y2}[c]{$\frac{\sigma\times {\rm
      BR}}{(\sigma\times
    \rm BR)_{\rm SM}}$}
\psfrag{y}[c]{${\rm BR}\qquad~ \frac{m_{\chi}}{100\,{\rm GeV}}$}
\includegraphics[width=0.47\textwidth]{brM2_200mu200Mchi_35_mst_BG_Dark.eps}\qquad
\includegraphics[width=0.47\textwidth]{decM2_200mu200Mchi_35_mst_BG_Dark.eps}
\caption{Evolution along the path BG of Fig.~\ref{figure-1} of
  $\sigma\times {\rm BR}/(\sigma\times {\rm BR})_{SM}$ (left panels) and BR
  (right panels) for $M_2=200$ GeV. $M_1$ and $\mu$ are fixed such
  that $m_{\chi^0_1}=35$\,GeV and $g_{Z11}=0.05$.}
\label{figure-11}
\end{center}
\end{figure}

Summarizing once one fixes $M_2=200\,$GeV, for $40.6\,{\rm GeV}\leq
M_1\leq 55$\,GeV and $2.4\leq\tan\beta\leq 15$, it is possible to
properly choose $\mu$ in order to obtain the correct DM density and
satisfy all experimental as well as EWBG constraints. Since the points
in Fig.~\ref{figure-1} are roughly independent of $M_1,\, M_2$ and
$\mu$, the interval in $\tan\beta$ parametrizes univocally the BG path
if one imposes $m_h=125.5\,$GeV (as well as $m_Q=10^6\,$TeV and
$v(T_n)\simeq T_n$). Using this parametrization we plot the production
cross-sections and Higgs branching ratios along the BG path in
Fig.~\ref{figure-11}. As compared with Fig.~\ref{figure-10} the
production cross-sections are smaller even though in the latter the
neutralino is lighter than 35 GeV.  This is due to the smaller values
of $\mu$ that are used to achieve $g_{Z11}=0.05$~\footnote{Indeed, as
  stressed in section~\ref{correlationsneut} the production
  cross-sections rates presented in previous sections can all be
  increased or decreased by moving $\mu$ and $M_2$ away from the
  considered values, $\mu=200\,$GeV and $M_2=200\,$GeV. }. The best
agreement with LHC data corresponds to values of the stop mass
$m_{\tilde t}\simeq 105$ GeV.

For neutralino masses larger than 35-40\,GeV (but still smaller than
$m_Z/2$), and/or $g_{Z11}>0.05$ the neutralinos would yield too small
thermal relic density today and they could just be a component of DM
in the Universe.  On the other hand, for masses lighter than 35-40 GeV
and/or $g_{Z11}< 0.05$, if the lightest neutralinos were the LSP they
would overclose the Universe. There should exist therefore a lighter
supersymmetric particle into which the neutralino could decay. The
possible candidates would be either light gravitinos or axinos.

A light enough gravitino leading to neutralino decays at times
previous to the nucleosynthesis era would be an obvious choice that
would lead to no cosmological constraints. Moreover one could demand
this lifetime to be large enough not to affect the collider
constraints. The reheating temperature should be small enough to avoid
overproduction of gravitinos in the early Universe, but larger than
$\mathcal O(100$\,GeV) in order to allow the mechanism of EWBG.

For small values of the gravitino mass the thermal gravitino relic density is given by~\cite{Pradler:2006hh}
\begin{equation}
\Omega_{3/2} h^2 \simeq 0.5 \; \left(\frac{M_{\tilde{g}}}{1 \; {\rm TeV}}\right)^2\frac{100 \; {\rm MeV}}{m_{3/2}}\frac{T_R}{10^{6} \; {\rm GeV}}~,
\end{equation}
where $M_{\tilde{g}}$ is the gluino mass and $T_R$ the reheating temperature. 
Therefore for gravitino masses of order of a few tens of MeV,
reheating temperatures smaller than about $10^5$~GeV are necessary in
order to recover agreement with the observed DM density.  On
the other hand, for a bino-like neutralino, the neutralino width
decaying into a photon and a gravitino is given by
\begin{equation}
\tau_{\chi^0_1} \simeq 8 \times 10^{7}~{\rm s} \; \left(\frac{10~{\rm GeV}}{m_{\chi^0_1}}\right)^5 \left(\frac{m_{3/2}}{100~{\rm MeV}} \right)^2~.
\end{equation}
For light stops, and assuming all other squarks are heavy, the current
bounds on the gluino mass are of order~TeV.  As a
possible working example, one can consider $M_{\tilde{g}} \simeq
1.4$~TeV, $T_R \simeq 1$~TeV, $m_{3/2} \simeq 1$ MeV and
neutralino masses of order 40-50~GeV, for which one obtains the proper
relic density with a neutralino decaying into gravitinos with a
lifetime smaller than a few seconds and therefore not
subject to any cosmological or astrophysical constraints.

Another viable candidate for thermal DM in the light
neutralino scenario we are considering is the axino $\tilde a$, the
fermionic supersymmetric partner of the axion $a$. As the MSSM does
not provide any solution to the strong CP problem the simplest
possibility is to add a (supersymmetrized) axion field, the Goldstone
boson of a global PQ symmetry broken at the scale $f_a$ ($5\times
10^9~\textrm{GeV}\lesssim f_a\lesssim 10^{11}~\textrm{GeV}$) that
solves  dynamically the strong $CP$ problem. Through the
supersymmetrization of the anomalous axionic coupling there appears
the Lagrangian term~\cite{Covi:2001nw}
\be
\mathcal L_{\tilde a}=\frac{g^{\prime\,2} C_a}{32\pi^2 f_a}\bar{\tilde a}\gamma_5 \sigma^{\mu\nu}\tilde B B_{\mu\nu}~,
\label{axinocoup}
\ee
where $C_a$ is an $\mathcal O(1)$ model dependent constant, and which
generates in particular the decay $\chi_1^0\to\tilde
a\gamma$. Neutralinos should decay at times previous to the
nucleosynthesis era not to spoil the BBN predictions. On the other
hand, as in the case of gravitino DM, the reheating
temperature should be small enough to avoid overproduction but not
below the electroweak phase transition temperature to allow the
EWBG mechanism to work.

The thermal axino relic density is given by~\cite{Brandenburg:2004du}
\be
\Omega_{\tilde a}h^2\simeq 0.12 \left( \frac{m_{\tilde a}}{5~\textrm{MeV}}\right)\,
\left( \frac{10^{10}~\textrm{GeV}}{f_a}\right)^2\,
\left( \frac{T_R}{10^3~\textrm{GeV}}\right)~,
\label{omegaaxino}
\ee
which is normalized to the observed value and shows the range of
parameters, $m_{\tilde a}\lesssim$ MeV~\footnote{It has been observed
  that the axino mass can be hierarchically smaller than the gravitino
  mass~\cite{Chun:1992zk}.} and $T_R\gtrsim $ TeV, leading to the
correct relic density.  On the other hand the lifetime of neutralinos
for the decay channel $\chi_1^0\to\tilde a\gamma$ is given
by~\cite{Covi:2001nw}
\be \tau_{\chi_1^0}\simeq 0.41\, {\rm s}~ \frac{1}{C_a^2 N_{11}^2}\left(128
  \alpha_{EM}\right)^{-2}
\left(\frac{20~\textrm{GeV}}{m_{\chi_1^0}}\right)^3\left(\frac{f_a}{10^{10}~\textrm{GeV}}\right)^2~,
\ee
where $m_{\tilde a}\ll m_{\chi_1^0}$ is assumed and $N_{11}$ is the
projection of $\chi_1^0$ along the Bino (for our choice of parameters
it is $N_{11}\sim 1$). Then the BBN bound $\tau_{\chi_1^0}\lesssim
1\,s$ is easily evaded for neutralino masses in the ballpark of
20 GeV.

\section{\sc Conclusions and outlook}
\label{conclusions}

In this article we have re-analyzed the LHC constraints on the Light
Stop Scenario, a framework of the MSSM where stops lighter than
120\,GeV are required by successful EWBG~\cite{Carena:2008rt,
  Carena:2008vj}.  In this scenario the gluon fusion production rate
tends to be enhanced by more than fifty percent with respect to the SM
rate, while the width of the Higgs decay into vector gauge bosons, as
well as into quarks and leptons, tends to be close to the SM
rate. Moreover the Higgs diphoton decay width tends to be somewhat
smaller than the SM one.

As previously stated in Refs.~\cite{Cohen:2012zz, Curtin:2012aa}, for
$m_h\simeq 125\,$GeV there is some tension between the recent Higgs
search results at the LHC and the predictions of the Light Stop
Scenario, putting strong constraints on the realization of this
scenario. These constraints do not only depend on the spectrum of the
Light Stop Scenario but also on the specific values of the couplings
obtained by requiring EWBG in the MSSM. Taking into account this fact
we have found smaller deviations from the Standard Model than those
determined in Ref.~\cite{Curtin:2012aa}.  Moreover we have highlighted
that much better agreement with LHC results can be achieved if the
lightest neutralino mass is smaller than about 60\,GeV. For such light
neutralinos the Higgs may have a significant invisible decay width,
which may substantially modify the Higgs branching ratios into SM
particles.

The precise prediction of the Higgs signatures of the Light Stop
Scenario in the presence of light neutralinos does not only depend on
the neutralino mass, which is controlled by the bino mass parameter
$M_1$, but on $\tan\beta$ and the Higgsino mass parameter
$\mu$. Larger values of $\mu$ and/or $\tan\beta$ lead to a suppression
of the coupling of the neutralino to the SM-like Higgs and therefore
to smaller neutralino effects.  Branching ratios of the Higgs decaying
into neutralinos of order 30--60\% lead to a good agreement of the Light
Stop Scenario predictions with the LHC Higgs data. These branching
ratios may be obtained for values of the neutralino mass close to
$m_h/2$ for low values of $\mu$ and $\tan\beta$ or for larger values
of $\mu$ and $\tan\beta$ in the case of light neutralinos, $m_{\chi_1^0} < 45$~GeV.

There are some general features that characterize the proposed
scenario:
\begin{itemize}
\item  The Higgs production channels coming from gluon fusion must have
enhanced rates with respect to the SM ones.
\item The Higgs production channels coming from weak boson fusion must
have suppressed rates with respect to the SM ones. 
\item The $h \to ZZ$ and $h \to WW$ decay rates should be slightly larger
than the $h \to \gamma \gamma$ rate in both the gluon fusion and vector boson fusion
production channels. 
\item Apart from a small variation induced by the change in the $h \to
  \gamma\gamma$ width, for a given Higgs mass the ratios between the
  different decay channels coming from gluon fusion (or vector boson
  fusion) are roughly independent of the stop mass and these ratios
  are not changed by the Higgs decay into neutralinos.
\end{itemize}
These features are compatible with the present LHC data and they shall
be scrutinized with more precise measurements of the Higgs decay
rates.  In particular, a better understanding of the gluon fusion
contribution to the dijet channel is required in order to compare
measurements on this process with the vector boson fusion decay $h\to
\gamma \gamma$ predicted in the Light Stop Scenario, where the gluon
fusion Higgs production is enhanced with respect to the SM one.  Due
to these unknown systematic uncertainties and lack of statistics, a
detailed comparison with LHC data has been left for a future analysis.

Besides the measurements of the visible Higgs decay channels, further
issues that we have taken into account in the present analysis are:
\begin{itemize}
\item{\it Higgs invisible width:} The analysis we have performed is
  very sensitive to this constraint but at present large regions of the
  parameter space are still allowed. Indeed, current uncertainties on
  this decay channel are too large to put strong constraints on the model, but
  in the near future they are expected to substantially decrease. 
\item{\it Stop searches:} Light neutralinos can qualitatively modify
  the stop signatures at LHC. Depending on the neutralino mass the
  stop is expected to decay into three or four bodies, which finally
  appear as one jet, one or two leptons and missing energy. Contrarily
  to what was usually believed for the Light Stop Scenario, these
  many-body stop decay channels tend to dominate the two-body decay
  $\tilde t \rightarrow c\chi_1^0$.  The present status of stop
  searches put constraints that are strongly model dependent and, in
  general, do not rule out the considered scenario. However dedicated
  analyses would be worthwhile to probe the existence of light stops
  and neutralinos, in particular in the presence of light
  $\tilde\tau$'s or $\tilde{\nu}_{\tau}$'s at the electroweak scale or
  of several competitive stop decay channels.
\item{\it $Z$ invisible width:} Light neutralinos increase the $Z$
  invisible decay width. The SM itself predicts the width of the Z boson
  into neutrinos to be  about one sigma above the LEP
  measurement. In our analysis we have constrained the $Z$ invisible
  width to be compatible with the LEP bound at around 95\%
  C.L.\,.
\end{itemize}

Further phenomenological issues can be considered besides collider
physics. The quest for DM candidate is one of them. We have described
different solutions to this puzzle within the proposed scenario. For
instance if the lightest neutralino is the LSP and has a mass
$m_{\chi_1^0}=35\div40$ GeV, its thermal relic abundance is mainly
determined by its coupling to the $Z$ boson. In part of the parameter
space such neutralinos provide the correct DM density, allowing in
addition for sufficiently-large Higgs invisible decay. Alternatively,
one can also assume that the neutralino is not the LSP. This opens up
a wide choice of DM frameworks. For instance, for reheating
temperatures of the order of $10\div100$ times the electroweak scale,
either gravitinos or axinos are plausible DM candidates.

To conclude,  the LSS scenario for EWBG in the MSSM is currently being
probed at the LHC.  Higgs and stop searches are already putting strong constraints
on the possible realization of this scenario and more relevant information should be 
gathered  at the end of the present year, when the total integrated
luminosity in the most sensitive Higgs search channels will be significantly  larger that the present one. 
More precise data on the Higgs production in the different
channels as well as further stop searches at the LHC will be of
paramount importance for the ultimate verdict on the feasibility of this scenario.

\subsection*{\sc Acknowledgments}
\noindent 
GN thanks F.~Recchia and A.~Triossi for programming
troubleshooting. MQ was supported in part by the Spanish
Consolider-Ingenio 2010 Programme CPAN (CSD2007-00042) and by
CICYT-FEDER-FPA2008-01430. Fermilab is operated by Fermi Research
Alliance, LLC under Contract No. DE-AC02-07CH11359 with the
U.S. Department of Energy. Work at ANL is supported in part by the
U.S. Department of Energy~(DOE), Div.~of HEP, Contract
DE-AC02-06CH11357.

\end{document}